\documentclass[%
 reprint,
 amsmath,amssymb,
 aps,prc,
 superscriptaddress,
]{revtex4-1}

\usepackage{multirow}
\usepackage{graphicx}
\usepackage{dcolumn}
\usepackage{bm}
\usepackage{hyperref}

\hypersetup{
    colorlinks=true,       
    linkcolor=black,    
    citecolor=blue,        
    filecolor=magenta,      
    urlcolor=black           
}

\begin{document}


\title{Structure of $\textsuperscript{20}$Ne states in the resonance \textsuperscript{16}O+$\alpha$ elastic scattering}

\author{D. K. Nauruzbayev}
 \email{dosbol.ndk@gmail.com}
 \affiliation{National Laboratory Astana, Nazarbayev University, Astana, 010000, Kazakhstan}
 \affiliation{Saint Petersburg State University, Saint Petersburg, 199034, Russia}

\author{V. Z. Goldberg}
 \affiliation{Cyclotron Institute, Texas A$\&$M University, College station, Texas, 3366, USA}

\author{A. K. Nurmukhanbetova}
 \email{anurmukhanbetova@nu.edu.kz}
 \affiliation{National Laboratory Astana, Nazarbayev University, Astana, 010000, Kazakhstan}

\author{M. S. Golovkov}
 \affiliation{Joint Institute for Nuclear Research, Dubna, 141980, Russia}
 \affiliation{Dubna State University, Dubna, 141980, Russia}

\author{A. Volya}
 \affiliation{Department of Physics, Florida State University, Tallahassee, Florida, 32306, USA}

\author{G. V. Rogachev}
 \affiliation{Cyclotron Institute, Texas A$\&$M University, College station, Texas, 3366, USA}

\author{R. E. Tribble}
 \affiliation{Cyclotron Institute, Texas A$\&$M University, College station, Texas, 3366, USA}

\begin{abstract}
\begin{description}
\item[Background]
The nuclear structure of the cluster bands in \textsuperscript{20}Ne presents a challenge for different theoretical approaches. It is especially difficult to explain the broad 0$^+$, 2$^+$ states at 9 MeV excitation energy. Simultaneously, it is important to obtain more reliable experimental data for these levels in order to quantitatively assess the theoretical framework.
\item[Purpose]
To obtain new data on \textsuperscript{20}Ne $\alpha$ cluster structure.
\item[Method]
Thick target inverse kinematics technique was used to study the \textsuperscript{16}O+$\alpha$ resonance elastic scattering and the data were analyzed using an \textit{R} matrix approach. The \textsuperscript{20}Ne spectrum, the cluster and nucleon spectroscopic factors were calculated using cluster-nucleon configuration interaction model (CNCIM).
\item[Results]
We determined the parameters of the broad resonances in \textsuperscript{20}Ne: 0$^+$ level at 8.77 $\pm$ 0.150 MeV with a width of 750 (+500/-220) keV; 2$^+$ level at 8.75 $\pm$ 0.100 MeV with the width of 695 $\pm$ 120 keV; the width of 9.48 MeV level of 65 $\pm$ 20 keV and showed that 9.19 MeV, 2$^+$ level (if exists) should have width $\leq$ 10 keV.
The detailed comparison of the theoretical CNCIM predictions with the experimental data on cluster states was made.
\item[Conclusions]
Our experimental results by the TTIK method generally confirm the adopted data on $\alpha$ cluster levels in \textsuperscript{20}Ne. The CNCIM gives a good description of the \textsuperscript{20}Ne positive parity states up to an excitation energy of $\sim$ 7 MeV, predicting reasonably well the excitation energy of the states and their cluster and single particle properties. At higher excitations, the qualitative disagreement with the experimentally observed structure is evident, especially for broad resonances.

\end{description}
\end{abstract}

\pacs{Valid PACS appear here}
\maketitle


\section{\label{sec:level1}INTRODUCTION}

It is well recognized that the $\alpha$ particle interaction with atomic nuclei is important in astrophysics \cite{1}. Even if astrophysical reactions involving helium do not proceed through the strong $\alpha$-cluster states (because of their high excitation energy), these states can provide $\alpha$ width to the states that are closer to the region of astrophysical interest through configuration mixing.

For a long time, the surprising alpha cluster structure has been a stimulus for the development of classical shell model approaches (see \cite{2} for new results). Additionally, work by the authors of Ref. \cite{3} recently ``strengthened the theoretical motivation for experimental searches of alpha cluster states in alpha-like nuclei" \cite{3}. The authors of Ref. \cite{3} related the nuclear structure in light systems of even and equal numbers of protons and neutrons with the first-order transition at zero temperature from a Bose-condensed gas of alpha particles (\textsuperscript{4}He nuclei) to a nuclear liquid.

\textsuperscript{20}Ne nucleus presents a famous example of the manifestation of the alpha-cluster structure, and therefore it makes this nucleus a touchstone for \textit{ab initio} approaches \cite{4}. The nucleus \textsuperscript{20}Ne is a benchmark case for the traditional shell model and its extension into algebraic and clustering domains. The well-established effective interaction Hamiltonians such as \cite{5} not only shows an outstanding agreement with experimental data for \textit{sd}-shell nuclei but also generates configuration mixing that shows transition to deformation and clustering.

The remarkable feature of \textsuperscript{20}Ne nucleus shows up in the fact that almost all the observed states below 10 MeV can be classified into several overlapping rotational-like bands with the first one based on the ground  0$^+_1$ state. There are three other bands based on 0$^+$ levels: on 0$^+_2$ at 6.725 MeV, on a very narrow 0$^+_3$ at 7.191 MeV, on a very broad 0$^+_4$ at $\sim$ 8.7 MeV \cite{6} which are of evident cluster structure. The 0$^+_2$ and 0$^+_4$ bands have $\alpha$+\textsuperscript{16}O core structure as can be seen from their reduced $\alpha$ particles widths, and probably the 0$^+_3$ band has predominant \textsuperscript{12}C+\textsuperscript{8}Be structure which manifests used itself in the selectivity of the \textsuperscript{8}Be transfer reactions \cite{7}. As the ground state band and the 0$^+_2$ band in \textsuperscript{20}Ne can be related with similar structures in \textsuperscript{16}O and \textsuperscript{12}C, the ``additional" structure of 0$^+_4$ states is not understood \cite{4}. The cluster approaches \cite{8} related the 0$^+_4$ band with large $\alpha$-widths, starting with the so-called ``\textsuperscript{16}O+$\alpha$" higher nodal band, which has one more nodal point in ``\textsuperscript{16}O+$\alpha$" relative wave function than the lower bands have. However it appeared that there are too many bands with a similar structure.

The $\alpha$ particle decay threshold in \textsuperscript{20}Ne is 4.73 MeV, while the threshold for proton decay is at 12.8 MeV (neutron decay threshold is even higher). Therefore resonance $\alpha$ particle scattering should be considered as an evident way to obtain data on the natural parity levels in \textsuperscript{20}Ne up to 13 MeV excitation energy. Indeed the majority of adopted data \cite{6} on level properties in the region in question based on a resonance work and an analysis made in 1960 year. More recently the $\alpha$+\textsuperscript{16}O resonance scattering experiment were developed further to backward angles and the data were reanalyzed using \textit{R} matrix code Multi 6 \cite{9}. The authors \cite{9} obtained quite different results from those used in for many levels (see Table 1); in particular the broad 0$^+_4$ and 2$^+_4$ levels appeared even much more broader. The authors \cite{9} noted difficulties of the fit in the region of 6-8 MeV excitation energies, in a region mainly free from narrow resonances. Evidently strong states of over 1 MeV width should influence a very broad excitation region (see for instance \cite{10}).

T. Fortune et al. \cite{11,12} were the first who recognized the importance of the fact that single particle structure of the broad states is in drastic contradiction with the shell model predictions. They \cite{11,12} also proposed the idea of mixing different configurations to explain the effect. The same idea was used in \cite{13}. However, the authors of Refs. \cite{11,12} used old data for the 0$^+_2$ state and used some estimates for the properties of the broad states proposed in Ref. \cite{14} to support the idea. Later measurements \cite{6} gave the width of 19 keV for the 6.72 MeV state (which is about 25\% larger than that used in work \cite{11}).

\begin{figure}[!t]
    \begin{center}
    \include{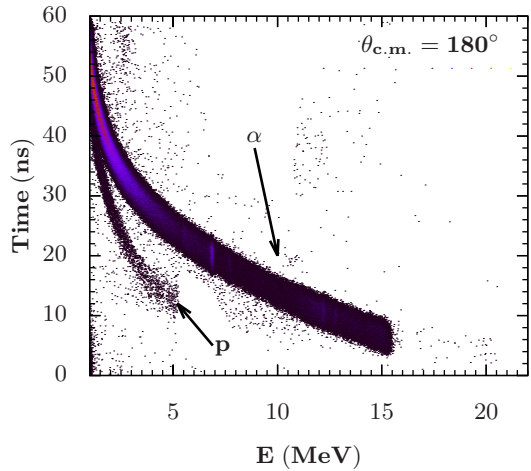}
    \end{center}
    \caption{E-T spectrum for the zero degrees detector. Alpha particles dominate; one can see also a weaker proton locus below the alpha particles.}
\end{figure}

The experimental aim of this work is to obtain new information on the structure of \textsuperscript{20}Ne states, especially the broad 0$^+$, 2$^+$ states. Unlike other experimentalists, we used the Thick Target Inverse Kinematics method (TTIK) (see \cite{15,16,17,18,19} and references therein) to study the excitation functions for the \textsuperscript{16}O($\alpha$, $\alpha$)\textsuperscript{16}O elastic scattering in the \textsuperscript{20}Ne excitation region of 5.5-9.6 MeV and in a broad angular interval. On the theoretical side, we also used multi configuration shell model calculation to understand the limits of this approach in a description of the cluster states.

\section{\label{sec:level2}EXPERIMENT}

\begin{figure}[!t]
    \begin{center}
    \include{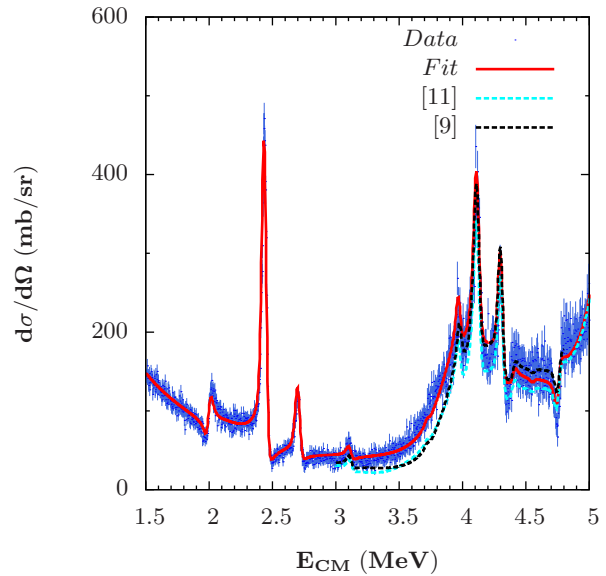}
    \end{center}
    \caption{The \textsuperscript{16}O($\alpha$, $\alpha$)\textsuperscript{16}O elastic scattering excitation function  at 180$^\circ$ cm. The excitation energies in \textsuperscript{20}Ne, E$_x$ in Table 1,  are related with cm energy, E$_{cm}$, by expression,  E$_x$=E$_{cm}$+4.73 MeV.  The bold (red) line is the \textit{R} matrix fit with the parameters of the present work. The dot (cyan) line is a fit with the 0$^+_4$ excitation energy of 8.3 MeV \cite{11}, and dot (black) line is a fit with the 0$^+_4$ energy excitation energy of 8.62 MeV and the width of 1.472MeV \cite{9}.}
\end{figure}

The experiment was performed at the DC-60 cyclotron (Astana) \cite{17} which can accelerate heavy ions up to the 1.9 MeV/A energy. While the TTIK method can't compete with a classical approach in terms of energy resolution, the possibility of observing excitation functions at and close to 180 degrees, where the resonance scattering dominates over the potential scattering, enables one to obtain more reliable information on the broad states.  In the TTIK technique the inverse kinematics is used; and the incoming ions are slowed in a helium target gas. The light recoils, $\alpha$ particles, are detected from a scattering event. These recoils emerge from the interaction with the beam ions and hit a Si detector array located at forward angles while the beam ions are stopped in the gas, as $\alpha$ particles have smaller energy losses than the scattered ions. The TTIK approach provides a continuous excitation function as a result of the slowing down of the beam.

For the present experiment, the scattering chamber was filled with helium of 99.99\% purity. The 30 MeV \textsuperscript{16}O beam entered the scattering chamber through a thin entrance window made of 2.0 $\mu$m Ti foil. Eight monitor Si detectors were placed in the chamber to detect \textsuperscript{16}O ions elastically scattered from the Ti foil at 21$^\circ$ angle. This array monitors the intensity of the beam with precision better than 4\%. Fifteen 10x10 mm$^2$ Si detectors were placed at a distance of $\sim$ 500 mm from the entrance window in the forward hemisphere at different laboratory angles starting from zero degrees. The gas pressure was chosen to stop the beam at distance of 40 mm from the zero degrees detector. The detector energy calibration and resolution ($\sim$ 30 keV) were tested with a \textsuperscript{226}Ra, \textsuperscript{222}Rn, \textsuperscript{218}Po and \textsuperscript{214}Po $\alpha$-source. The experimental set up was similar to that used before \cite{19}, and more details can be found in Ref. \cite{17,19}. The main errors in the present experimental approach are related to the uncertainties of the beam energy loss in the gas. To test the energy loss, we placed a thin Ti foil (2.0 $\mu$m) at different distances from the entrance window. This can be used during the experiment without cycling vacuum. We found that the data \cite{20} for energy loss of \textsuperscript{16}O in helium are correct. The details of these tests will be published elsewhere. As a result, we estimated that the uncertainties in the absolute cross section are less than 6\%. This conclusion was tested by comparison with the Rutherford cross sections at low energies. The agreement with the Rutherford scattering is within 5\% error bars at all angles (see Fig.3).

Together with the amplitude signal, the Si detectors provided for a fast signal. This signal together with a ``start" signal from RF of the cyclotron was used for the Time-of-Flight measurements. This E-T combination is used for particle identification in the TTIK approach \cite{16,17,18}. Of course, only $\alpha$ particles should be detected as a result of the interaction of \textsuperscript{16}O with helium at the chosen conditions. However, protons can be created in the Ti window, and protons can appear due to hydrogen admixtures in the gas. Indeed, we have observed a weak proton banana, likely as a result of  reactions in the window. These protons were easily identified by TF and separated from the $\alpha$ particles, as  seen in Fig.1.

\section{\label{sec:level3}Experimental Results and Discussion}

\begin{figure}[!t]
    \begin{center}
    \include{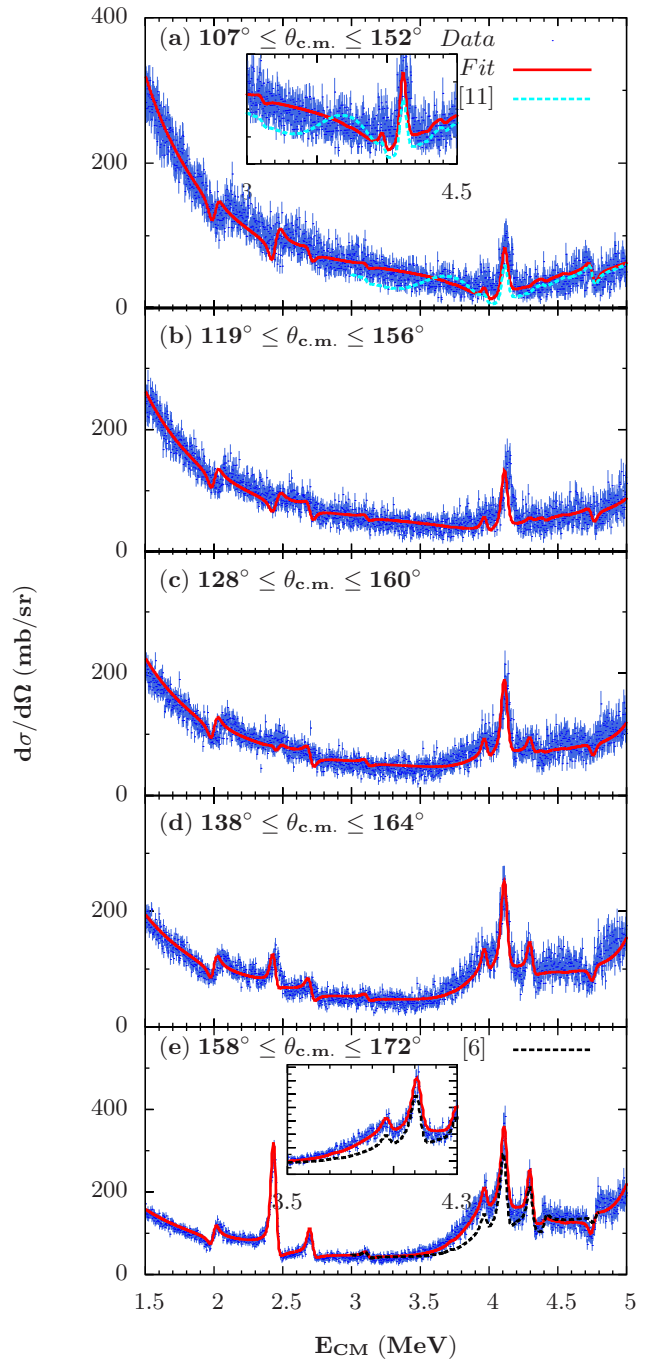}
    \caption{\textit{R} matrix fit (bold red curve) of the excitation functions for the $\alpha$+\textsuperscript{16}O elastic scattering. (a) The dashed (cyan) curve presents the data of  the 0$^+$ level at the 8.3 MeV excitation energy \cite{11} and (e) dot (black) line is a fit with 2$^+$ level at the excitation energy of 9.0 MeV \cite{6}.}
    \end{center}
\end{figure}   

\begin{table*}[!t]
\label{tab:1}
\caption{\textsuperscript{20}Ne levels}
\begin{center}
\begin{tabular}{lccccccccccclccl}
\hline
\hline
\multirow{1}{*}{N} & \multicolumn{3}{c}{TUNL data \cite{6}} & \multirow{1}{*}{} & %
    \multicolumn{2}{c}{H. Shen et al. \cite{9}} & \multirow{1}{*}{} & %
    \multicolumn{3}{c}{This work} & \multirow{1}{*}{} & \multicolumn{4}{c}{CNCIM} \\
\cline{2-4}\cline{6-7}\cline{9-11}\cline{13-16}
 & E$_x$ & J$^\pi$ & $\Gamma_\alpha$ & & E$_x$ & $\Gamma_\alpha$ & & E$_x$ &
 $\Gamma_\alpha$ & $\gamma_\alpha$ & & E$_x$ & J$^\pi$ & SF$_p$ & SF$_\alpha$\\
 & (MeV) & & (keV) & & (MeV) & (keV) & & (MeV) & (keV) & & & (MeV) &  &  & \\
\hline

1 &	0 & 0$^+_1$ &  & & - & - & & 0 & & Large & & 0 & 0$^+$ & 0.36 & 0.73\\
2 &	1.63 & 2$^+_1$ & &  & - & - & & 1.63 & & Large & & 2.242 & 2$^+$ & 0.41 & 0.67\\
3 &	4.25 & 4$^+_1$ & &  & - & - & & 4.25 & & Large & & 4.58 & 4$^+$ & & 0.62\\

4 &	5.78 & 1$^-$ & (28$\pm$3)x10$^{-3}$ &  & - & - &  & 4.45 & 0.03 & 1.4\\				
5 &	6.73 & 0$^+_2$ & 19$\pm$0.9 & & 6.72 & 11 & & 6.78 & 20.6 & 0.47 & & 6.94 & 0$^+_3$ & 0.55 & 0.46\\
6 &	7.16 & 3$^-$ & 8.2$\pm$0.3 & & 7.16 & 10 & & 7.18 & 8.3 & 1.37\\				
7 &	7.19 & 0$^+_3$ & 3.4$\pm$0.2 & & 7.19 & 5 & & 7.20 & 3	& 0.019 & & 6.27** & 0$^+_2$ & 0.055 & 0.44**\\
8 &	7.42 & 2$^+_2$ & 15.1$\pm$0.7 & & 7.43 & 7 & & 7.44 & 14.3 & 0.19 & & 7.39 & 2$^+_3$ & 0.01 & 0.12\\
9 &	7.83 & 2$^+_3$ & 2 & & 7.83 & 1 & & 7.85 & 3.68 & 0.01 & & 7.15** & 2$^+_2$ & 0.12 & 0.18**\\

10 & 8.45 & 5$^-$ & 0.013$\pm$0.004 & & 8.45 & 0.02 & & 8.45 & 0.013\\					
11 & 8.71 & 1$^-$ & 2.1$\pm$0.8 & & & & & 8.71 & 3.5\\					
12 & $\approx$8.7 & 0$^+_4$ & $>$800 & & 8.62 & 1470 & & 8.77$\pm$0.15 & 750$\pm$220 & $\sim$0.25 & & 9.66** & 0$^+_4$ & 0.002 & 0.18**\\
13 & 8.78 & 6$^+_1$ & 0.11$\pm$0.02 & & - & - & & 8.78 & 0.14 & 0.5 & & 9.49 & 6$^+$ & & 0.51\\
14 & 8.85 & 1$^-$ & 19 & & 8.84 & 27 & & 8.85 & 18.0\\					
15 & 9.00 & 2$^+_4$ & $\approx$800 & & 8.87 & 1250 & & 8.79$\pm$0.10 & 695$\pm$120 & 0.86 & & 8.36** & 2$^+_4$ & 0.02 & 0.02**\\
16 & 9.03 & 4$^+_3$ & 3 & & 9.02 & & & 9.03 & 1.9 & 0.03 & & 9.0 & 4$^+$ & & 0.09\\

17 & 9.12 & 3$^-$ & 3.2 & & 9.09 & 4 & & 9.13 & 4.1\\
18 & 9.19 & 2$^+$ &  &  & - & - & & (9.29) & $\leq$10	\\				
19 & 9.48 & 2$^+$ & 29$\pm$15 & & 9.48 & 46 & & 9.48 & 65$\pm$20 & 0.02?\\				
20 & 9.99 & 4$^+_4$ & 155$\pm$30 & & 10.02 & 150 & & 9.97 & 157 & 0.38 & & 9.5 & 4$^+$ & & 0.009\\
21* & 10.26 & 5$^-$ & 145$\pm$40 & & 10.26 & 190 & & 10.26 &	& 1.9	\\			
22 & 10.41 & 3$^-$ & 80 & & 10.40 & 101 & & 10.41 & &	\\

23 & 10.58 & 2$^+$ & 24 & & 10.56 & 15 & & 10.58 & & & & 10.2 & 2+ & 0.005 & 0.04\\
24 & 10.80 & 4$^+_4$ & 350 & & 10.75 & 400 & & 10.80 & & & & 10.7 & 4$^+$ & & 0.04\\
25 & 10.97 & 0$^+_5$ & 580 & & 10.99 & 700 & & 10.97 & & & & 11.9 & 0$^+$	\\	
26 & 11.24 & 1$^-$ & 175 & & 11.19 & 85 & & 11.24 & &\\					
27 & 11.95 & 8$^+$ &  (3.5$\pm$1.0)x10$^{-2}$  & & & & & 11.95 & & 0.35 & & 11.50 & 8$^+$ & & 0.40\\
\hline
\hline
\multicolumn{16}{l}{ * For the levels with numbers 21-27 the parameters of the present fit were fixed as in \cite{6} } \\
\multicolumn{16}{l}{ ** Calculated in \textit{psd} space. SF is to the first excited state in \textsuperscript{16}O; SFs for the ground state in \textsuperscript{16}O are $\leq$ 0.1 } \\

\end{tabular}
\end{center}
\end{table*}

The experimental excitation functions were analyzed using multilevel multichannel \textit{R} matrix code \cite{21}. The calculated curves were convoluted with the experimental energy resolution. The experimental energy resolution was $\sim$ 30 keV at zero degrees and deteriorated up to $\sim$100 keV with angles estranging from zero degrees. We did not notice a deterioration of the energy resolution with the energy loss of the beam in the chamber. As it seen in Table 1 the excitation energies of the resonances of the present work agree with the adopted ones \cite{6} within 10-15 keV. This agreement is an evidence of our overall good energy calibrations and the correct account of the ion energy loss in helium in the present work. Fig.2 and Fig.3 give the experimental excitation functions together with the present \textit{R} matrix fit. Fig.2 demonstrates the data at 180$^\circ$ and illustrates the differences in the fits due to different parameters of the broad 0$^+_4$ resonance. The data on the resonances used in the present \textit{R}-matrix fit are summarized in the Table 1 together with the adopted data \cite{6}. Data of the last \textit{R} matrix analysis \cite{9} are also given in Table 1. The analysis \cite{9} resulted in level parameters which are often different from the adopted.

Our analysis (Table 1) resulted in  small discrepancies with the data \cite{6} in the detail description of the narrow states (widths less than 10 keV). The \textit{R}-matrix code \cite{21} is tuned for the TTIK measurements and for the analyses of states with a width of over 10 keV to accelerate automatic fit calculations. Therefore the small disagreements for the narrow states are not significant. We focused on the broader states and on the part of the excitation function changing slowly with energy and angle.

Our analysis indicates that all strong alpha cluster states at 5$\sim$6 MeV below or up the investigated excitation energy region can influence the \textit{R}-matrix fit. Therefore we included in the fit the \textsuperscript{20}Ne ground state, the first 2$^+$ and 4$^+$ states and 1$^-$ (5.67 MeV) state in the fit, even though  below the investigated region (see Table 1). Among these states, only the 1$^-$ state is above the $\alpha$ particle decay threshold in \textsuperscript{20}Ne; the reduced width of this state is known, and it is large. Shell model calculations also give very large spectroscopic factors for all members of the ground state band. Indeed, a good agreement needs large values of the corresponding amplitudes (over 0.7). Above the investigated excitation energy region, high spin $\alpha$-cluster resonances are mainly known. Each of these resonances (see Table 1) considered separately influences the fit, especially at 180$^\circ$. However, their joint influence is much weaker. This cancellation is due to different parities of the spins. Only the influence of the closest to the investigated region, the 4$^+$ (9.99 MeV) resonance, can be noticed. A somewhat better fit needs the width of this resonance to be slightly larger $\sim$160 keV (well within the quoted uncertainties, see Table 1). Parameters of all other resonances above the investigated region were fixed according to the data of Ref. \cite{6}. A good general fit ($\chi^2$=1.1) was reached in this way without any backward resonance inclusion.

\begin{table*}[!t]
    \caption{$\alpha$+\textsuperscript{12}C levels in \textsuperscript{16}O}
    \centering
    \begin{tabular}{ccccccccc}
    \hline
    \hline    
    \multicolumn{1}{c}{\textsuperscript{16}O level} & \multirow{1}{*}{} & \multicolumn{1}{c}{$\Gamma_\alpha$ $_{exp}$ keV \cite{6} } & \multirow{1}{*}{} & \multicolumn{1}{c}{$\gamma_\alpha$(1); -V$_0$ MeV} & \multirow{1}{*}{} & \multicolumn{1}{c}{$\gamma_\alpha$(2); -V$_0$ MeV} & \multirow{1}{*}{} & \multicolumn{1}{c}{$\gamma_\alpha$(3); -V$_0$ MeV}\\ 
    \cline{1-1}\cline{3-3}\cline{5-5}\cline{7-7}\cline{9-9}
    1$^-$; 9.58 MeV & & 420$\pm$20 & & 0.70; 138.2 & & 0.72; 150.0 & & 0.84; 158.5 \\
    4$^+$; 10.36 MeV & & 26$\pm$3 & & 0.68; 125.3 & & 0.88; 139,6 & & 1.21; 143.2 \\
    \hline
    \hline    
    \end{tabular}
    \label{tab:2}
\end{table*}

All resonances are at the maximum at 180$^{\circ}$. The broad hump at this angle at cm energy of 4 MeV (Fig. 2) is a clear indication for the presence of low spin states, 0$^+$ and 2$^+$. Higher spin states are narrower. A single level (2$^+$) cannot produce the strong peak at different angles, and levels of different parity, such as 2$^+$ and 1$^-$, interfere destructively at 180$^{\circ}$. The present analysis resulted in two practically degenerate states at $\sim$ 8.8 MeV with the same width (see Table 1). Our results for the broad 2$^+$ are rather close to the adopted values. However, if this level is moved to 9.0 MeV excitation energy (as in \cite{6}) then the fit becomes worse, especially in the vicinity of the dip die to the presence of another 2$^+$ level at 9.48 MeV (Fig. 3(e)).  The fit becomes even worse with a very broad 2$^+$ level resulting from the parameters of Ref. \cite{9}.

T. Fortune at al., \cite{11} observed a broad distribution with a center at 8.3 MeV excitation energy in \textsuperscript{20}Ne and related it with the 0$^+$ level. Fig. 2 presents \textit{R} matrix calculations with the 0$^+$ level to be moved to 8.3 MeV excitation energy. This move destroyed the good fit. Fig.2 shows also that a very broad 0$^+$ level of the fit \cite{9} destroys the agreement.

The 18$^{th}$ (2$^+$) level of \cite{6} is in the energy region of the present investigation. It was observed in a single work in a study of \textsuperscript{20}Na $\beta^+$ decay. We have not found any reliable evidence for the presence of this state. If it exists, its width should be less than 10 keV. We observe a fluctuation of points which might be associated with a narrow 2$^+$ level at excitation energy of 9.29 MeV.

We noted that the 19$^{th}$ level, 2$^+$, has the adopted \cite{6} excitation energy in our fit but a different width of 65$\pm$20 keV. While the two times difference with Ref. \cite{6} in the widths is marginally exceeds the error bars, the influence on the better fit is evident. The adopted data \cite{6} for this level are based on the results of a single older work \cite{22}. The authors \cite{22} observed a weak $\gamma$ decay of this state in the presence of a large background. A broader level than in Ref. \cite{22} was also found in work \cite{9} as shown in Table 1.

We characterized the alpha-cluster properties of the states above the alpha particle decay threshold by SF=$\gamma_\alpha$=$\Gamma_\alpha$ $_{exp}$/$\Gamma_\alpha$ $_{calc}$, where $\Gamma_\alpha$ $_{calc}$  is the single alpha particle width calculated in the $\alpha$-core potential. To normalize SFs we calculated these values for the well-known alpha-cluster states in \textsuperscript{16}O.

The Woods-Saxon potential was used to calculate the limit ($\Gamma_\alpha$ $_{cal}$) as the widths of single particle states in the potential. First we tried to fit the widths of known alpha-cluster states, 1$^-$ and 4$^+$, in \textsuperscript{16}O so that $\gamma_\alpha$ = $\Gamma_\alpha$ $_{exp}$/$\Gamma_\alpha$ $_{cal}$ $\sim$ 1.0. The real part of the potential was changed to fit the binding energy of the states. The radius of the potential was chosen to be R = r$_0$ $\times$12$^{1/3}$; the Coulomb potential was taken in to account as a charge sphere potential with R$_{coul}$ = R.  We made first calculations (1) with r$_0$ = 1.31 fm and the diffuseness a = 0.65 fm, then we set r$_0$ = 1.23 fm (2), and we finally performed the third calculations (3)  with r$_0$ = 1.23 fm and a = 0.6 fm. The results are summarized in Table 2. The $\gamma_\alpha$ calculations for the \textsuperscript{20}Ne states were made with the final (3) parameters  should be compared with SF given by a theory.

\section{\label{sec:level4}Theoretical description of the $\alpha$ cluster states in \textsuperscript{20}$\textbf{Ne}$}

CNCIM \cite{2} is among the latest developments of the classical shell model approaches towards clustering. This model targets a combination of classical configuration interaction techniques with algebraic methods that emerge in the description of clustering.  The ability to construct a fully normalized set of orthogonal cluster channels is at the core of this approach; the overlaps of the shell model states with these channels are associated with spectroscopic factors and compared in Table 1 with the reduced widths obtained earlier. The CNCIM allows us to study clustering features that emerge in models with well-established traditional shell model Hamiltonians. These effective model Hamiltonians are built from fundamental nucleon-nucleon interactions followed by phenomenological adjustments to select observables; thus, they generally describe of a broad scope of experimental data with high accuracy. Apart from using these phenomenological shell model Hamiltonians, our study does not involve any adjustable parameters. In order to fully explore the problem, we considered several different model spaces and corresponding Hamiltonians: the \textit{sd} model space with USDB interaction \cite{5}; unrestricted \textit{p-sd} shell model Hamiltonian \cite{23}, the same Hamiltonian has been used in Refs. \cite{2};  WBP Hamiltonian \cite{24} allowing 0$\hbar\omega$, 1$\hbar\omega$  and 2$\hbar\omega$ excitations in \textit{p-sd-pf} valence space; and the \textit{sd-pf} Hamiltonian \cite{25}. This sequence of Hamiltonians represents and expansions of the valence space from \textit{sd} to \textit{p-sd}, to \textit{p-sd-pf}. All models are in good agreement for the \textit{sd}-states; the low-lying negative parity states as well as positive 2ph excitations are dominated by the \textit{p-sd} configurations. Thus, in Table 1 we only include the results from the \textit{p-sd} Hamiltonian which turned out to be most representative although the following discussion and conclusions are largely based on comparisons. The lowest states associated with significant \textit{fp} shell component appear at excitation energies above 15 MeV.

Shell model calculations for \textsuperscript{20}Ne with open \textit{2s-1d (sd)} shells predict well the ground state band. The structure of this band is based on the dominating SU(3) configuration (about 75\%) with quantum numbers (8,0). The model predicts (Table 1) large SF for all members of the band based on the ground state in \textsuperscript{20}Ne. The 0$^+$, 2$^+$, and 4$^+$ members of the band are below the $\alpha$ particle decay threshold and do not have observable $\alpha$ particle widths. However, large $\alpha$ cluster SFs for these states provide for a better \textit{R} matrix fit. While uncertainties for the \textit{R} matrix amplitudes for these states are large, the fit (Fig. 2 and 3) requires these amplitudes to be close to those of the negative parity states with the known extreme $\alpha$ cluster structure. The $\alpha$ particle widths of the highest 6$^+$ and 8$^+$ members of this band are known. There is a long history of attempts and ideas to describe these widths using shell model approaches (see, for instance \cite{5}). The most calculations predicted large clustering for the band but could not explain the decrease of the reduced width for the 6$^+$ and 8$^+$ members. As one sees in Table 1, the CNCIM calculations are in fine agreement with the experimental data for these states. All members of this band have significant clustering that diminishes at higher energies due to configuration mixing.  The second 0$^+$ state within \textit{sd} space appears at around 6.7 MeV of excitation (in \textit{psd} model in Table 1, this is a third 0$^+$ state at 6.9 MeV). This level also has a substantial clustering component and absorbs nearly all 15\% of the remaining strength of the SU(3) (8,0) component. The following 2$^+_2$ (in experiment and in \textit{sd} model, but third in \textit{p-sd} model as discussed in what follows) at about 7.4 MeV of excitation energy being a member of the 0$^+_2$ band can be described in a similar way.

A defrost of the \textit{1p} shell (which is filled in \textsuperscript{16}O) results in the doubling of the levels, as it is observed and is shown in Table 1 new levels 0$^+$ and 2$^+$, marked with asterisks, appear. In our calculations, the ordering in energy is reversed for both doublets. Structurally the members of each doublet are very different which allows them to be so close in energy and inhibits configuration mixing and Wigner repulsion. One of the doublet levels (the lower 0$^+$ and 2$^+$ levels) has a large $\alpha$ cluster SF relative to the first excited state in \textsuperscript{16}O and much smaller SF relative to the ground state in \textsuperscript{16}O (pay attention that the theory gives the wrong order for the levels in question, see Table 1). Indeed the predicted difference in the structure is supported by population selectivity in different nuclear reactions. The 6.72 MeV 0$^+$ and 7.42 MeV 2$^+$ are populated much stronger than neighboring 7.20 and 7.83 MeV levels in the \textsuperscript{16}O(\textsuperscript{6}Li, d) reaction \cite{26}. The opposite is the case in the \textsuperscript{12}C(\textsuperscript{9}Be, n) or \textsuperscript{12}C(\textsuperscript{12}C,$\alpha$) \cite{27} reactions. 

The theory gives large single particle spectroscopic factors for the \textit{sd} states and smaller for the 7.20 and 7.83 MeV states. Indeed, one expects that states in \textsuperscript{20}Ne with a hole in the \textit{p1/2} shell will be weakly excited in the single nucleon transfer, \textsuperscript{19}F(\textsuperscript{3}He, d) reaction in accordance with the experimental data \cite{28}. The experiment \cite{11,12,28} supports also detailed single particle SF calculations giving SF for the ground state smaller than for the 0$^+$ 6.73 MeV state and much higher SF for the first excited 2$^+$ than for 2$^+$ member of the band based on the 6.73 MeV state.

In the \textit{sd} valence space (USDB) the third 0$^+$ state appears only at 11.9 MeV, and it has a relatively small alpha spectroscopic factor; opening of the \textit{p}-shall in addition to 0$^+$ at 6.27 MeV, leads to 0$^+$ state at 9.7 MeV. However, the predicted state has a low proton spectroscopic factor which is not as well supported by observations. A similar serious discrepancy is observed with a broad 4$^{th}$ 2$^+$ state at around 9 MeV, both \textit{sd} and \textit{p-sd} shell models produce candidates but with very low alpha SF.  As evident from our studies the only strong coupling to alpha channels could come from \textit{fp} shell and higher shells. The lowest two bands saturate the alpha strength within \textit{sd} configurations, holes in the \textit{p}-shell do not lead to a significant contribution due to low level of core excitation in the ground state of \textsuperscript{16}O. While our models predict high excitation energies of states with significant \textit{fp} components, we can speculate that strong configuration mixing, collective effects such as deformation, and coupling to continuum via super radiance mechanism \cite{29,30} can enhance admixture needed to reproduce the broad resonances observed. There is a similar problem with $\alpha$ cluster negative parity states 1$^-$ and 3$^-$, the \textit{p-sd} Hamiltonian produces an acceptable spectrum but the alpha spectroscopic factors are low (see also \cite{31}).

\section{\label{sec:level5}Conclusions}

In this work we study $\alpha$-clustering in \textsuperscript{20}Ne. This nucleus is a benchmark example of many theoretical techniques targeting clustering in light nuclei. Our \textit{R}-matrix analysis of TTIK experimental data confirms previously known results and establishes new constraints for the positions and widths of the resonances. We compared our findings with those obtained theoretically using cluster-nucleon configuration interaction approach developed in several previous works \cite{2} and references therein. There is good overall agreement between theoretically predicted and observed spectra. Our theoretical approach describes very well the ground state band and the band built on the first 0$^+$ state. Allowing cross shell excitations from the \textit{p}-shell, it was possible to reproduce the band built on the second 0$^+$ state. For these states, all spectroscopic factors for alpha transitions to the ground state of \textsuperscript{16}O and to the first excited state in \textsuperscript{16}O as well as proton spectroscopic factors to the ground state of \textsuperscript{19}F are well reproduced. The situation is not as good when it comes to resonances 1$^-$ 3$^-$ and 4$^{th}$ 0$^+$ and 4$^{th}$ 2$^+$, all of these states are broad and have exceptionally large alpha spectroscopic factors. 

In order to describe strong clustering features these states must include configurations from \textit{fp} shell and from higher oscillator shells, however Hamiltonians that we explored predict these contributions to be negligible below 15 MeV of excitation. Thus, inability of theoretical models to describe broad states exclusively while working well elsewhere suggests an additional coupling mechanism unaccounted for in the traditional shell model Hamiltonians. The super radiance suggested in Refs. \cite{29,30} could provide this mechanism. Alternatively, the problem could be associated with relatively unknown cross shell interactions. Therefore, work shows that the experimental study of alpha clustering represents an outstanding tool for exploring cross shell excitations especially those of multi-particle multi-hole nature.

\begin{acknowledgments}
This work was supported by Ministry of Education and Science of the Republic of Kazakhstan (grant number \#0115РК03029 ``NU-Berkeley", 2014-2018; grant number \#0115РК022465, 2015-2017). This material is also based upon work supported by the U.S. Department of Energy Office of Science, Office of Nuclear Physics under Grants No DE-FG02-93ER40773 and No. DE-SC0009883.  G.V.R is also grateful to the Welch Foundation (Grant No. A-1853).
\end{acknowledgments}

\end{document}